\documentclass[aps,prd,superscriptaddress,preprint,nofootinbib]{revtex4-1}

\usepackage{amsmath}
\usepackage{graphicx}

\DeclareMathOperator{\shuffle}{\ensuremath{\sqcup\!\sqcup}}
\def\ket#1{\langle #1\rangle}

\begin{document}

\title{All Two-Loop MHV Amplitudes in Multi-Regge Kinematics From Applied Symbology}

\author{Alexander Prygarin}
\affiliation{Department of Physics, Brown University, Providence, RI 02912, USA}
\author{Marcus Spradlin}
\affiliation{Department of Physics, Brown University, Providence, RI 02912, USA}
\author{Cristian Vergu}
\affiliation{ETH Z\"urich, Institut f\"ur Theoretische Physik, Wolfgang-Pauli-Str.~27, CH-8093 Z\"urich, Switzerland}
\author{Anastasia Volovich}
\affiliation{Department of Physics, Brown University, Providence, RI 02912, USA}

\begin{abstract}
Recent progress on scattering amplitudes has benefited from the mathematical
technology of symbols
for efficiently handling the types of polylogarithm functions which
frequently appear in multi-loop computations.
The symbol for all two-loop MHV amplitudes in planar SYM theory is known, but explicit
analytic formulas for the amplitudes
are hard to come by except in special limits where things simplify,
such as multi-Regge kinematics.
By applying symbology we obtain a formula for the leading behavior of the imaginary
part (the Mandelstam cut contribution) of this amplitude in multi-Regge kinematics for any number of gluons.
Our result predicts a simple recursive structure which agrees with a direct BFKL computation
carried out in a parallel publication.

\end{abstract}

\maketitle

\tableofcontents

\section{Introduction}

Scattering amplitudes in gauge theories are
complicated quantities even in relatively simple cases such
as planar ${\cal N} = 4$
supersymmetric Yang-Mills (SYM) theory
and despite the dramatic recent improvements
in our understanding of the mathematical structure of this theory.
Some of this complication is unavoidable, since they depend
non-trivially on many independent variables,
and necessarily do so in terms
of complicated functions:
at weak coupling they can be
expressed in terms of certain transcendental functions, and at strong
coupling they compute minimal areas in anti-de Sitter space with prescribed
boundary conditions (see Ref.~\cite{Alday:2007hr}).
Moreover at any coupling they apparently compute the expectation
value of polygonal Wilson loops with lightlike edges, when suitably
defined (see
Refs.~\cite{Alday:2007hr,Drummond:2007aua,Brandhuber:2007yx,Mason:2010yk,CaronHuot:2010ek,Belitsky:2011zm}).

Pioneering progress towards taming at least some of this complexity has
been made in Ref.~\cite{Goncharov:2010jf}
by the introduction to the physics literature of the notion of
the symbol of a generalized polylogarithm function (see
Ref.~\cite{Goncharov:2009}).  The symbol encapsulates
much of the physically relevant information about an amplitude while
simultaneously trivializing all of the functional identities
which render it nearly impossible to work with
polylogarithm functions directly.
In particular the application of symbol technology (or ``symbology'') enabled
the determination of a relatively
simple ``one-line'' analytic
formula for the two-loop 6-particle MHV amplitude in
Ref.~\cite{Goncharov:2010jf}
(which had been
evaluated numerically in
Refs.~\cite{Drummond:2007bm,Bern:2008ap,Drummond:2008aq}
and analytically
in terms of several pages of polylogarithm functions
thanks to the heroic effort of
Refs.~\cite{DelDuca:2009au,DelDuca:2010zg}).
Let us emphasize that symbology is a mathematical tool
not restricted to SYM theory; see for example Ref.~\cite{Manteuffel} for
a successful application to top quark pair production in QCD.

In recent work explicit results for the symbols of further amplitudes
in SYM theory
have started to accumulate in the literature, including
the two-loop MHV amplitudes for all $n$ in Ref.~\cite{CaronHuot:2011ky},
the three-loop 6-particle MHV amplitude in
Refs.~\cite{Dixon:2011pw,CaronHuot:2011kk}, and
the two-loop NMHV amplitudes for 6 and 7 particles respectively
in Refs.~\cite{Dixon:2011nj} and~\cite{CaronHuot:2011kk}.
Before proceeding
let us note that that new techniques such as those developed in
Refs.~\cite{CaronHuot:2011kk,Bullimore:2011kg}
seem to hold promise for generating much more data.

Unlike the case studied in Ref.~\cite{Goncharov:2010jf},
none of these amplitudes can be expressed in terms of the classical
polylogarithm functions only.  Despite this additional complexity
it remains a very interesting open problem to find fully
analytic formulas for these amplitudes in terms of generalized
polylogarithms (see Ref.~\cite{Duhr:2011zq} for a possible algorithm towards
this end).

Given this complexity we are led to study various limits in which the
answers simplify and then to hope that more general lessons can be extracted from them.  One way
to simplify the problem is to consider
the special case when the 4-momentum of each particle (or equivalently,
all edges of
the corresponding Wilson loop) lies in a common two-dimensional space.
This has enabled some very simple results both at weak
(see Refs.~\cite{DelDuca:2010zp,Heslop:2010kq,Heslop:2011hv,Gaiotto:2010fk})
and strong coupling (Ref.~\cite{Alday:2009ga}).

Another limit in which scattering amplitudes simplify is in
multi-Regge kinematics,
a type of high-energy limit in which some kinematic invariants become
much larger than others in a particular way described
below.  In the multi-Regge regime scattering
amplitudes are expressed as an expansion both in powers of the coupling
constant $g^2 N_c$
and in powers
of $\log(s/s_0)$ where $s$ is some large kinematic invariant.
The coefficients
of this double series expansion are functions of the remaining finite
kinematic invariants.
The interested reader may consult Ref.~\cite{Forshaw:1997dc} for a
pedagogical introduction.
We will be working
in the leading logarithm approximation, in which
the effective summation parameter $g^2 N_c \log(s/s_0)$ is of order
of unity.
The choice of $s_0$ is then immaterial, and
since all of the quantities we discuss will be dual conformally
invariant and so depend only on cross-ratios (see
Refs.~\cite{Drummond:2007cf,Drummond:2007au}) we will always
be able to choose to express
the expansion parameter as $\log (1-u)$ for some cross-ratio $u$ which
is approaching unity.

In the Euclidean region where all energy
invariants $s_{ij}$ are negative
the multi-Regge behavior of MHV amplitudes
is consistent with the BDS ansatz of Ref.~\cite{Bern:2005iz} to all
orders (see Refs.~\cite{Brower:2008nm,Brower:2008ia,DelDuca:2008jg}).
However it was pointed out in Ref.~\cite{Bartels:2008ce} that in other regions the BDS ansatz is violated due to
the presence of Mandelstam cuts.
The difference between the actual MHV amplitude and the BDS ansatz is called the remainder function.
In those other regions some energy invariants $s_{ij}$ become positive.
This requires an analytic continuation whose effect at the level
of cross-ratios is to multiply each one by some phase.
This analytic continuation reveals the contribution from the Mandelstam cuts,
which dominate over Regge poles in the remainder
function, and behave at $L$-loop order like
$\log^{L-1} (s_{ij}/s_0)$, or $\log^{L-1} (1-u)$ when properly assembled into
cross-ratios.

Let us note that the expansion of amplitudes in $\log (1-u)$ bears some
superficial resemblance to the
Wilson loop OPE
expansion which has been studied quite fruitfully in a number
of recent papers (see
Refs.~\cite{Alday:2010ku,Gaiotto:2010fk,Gaiotto:2011dt,Sever:2011pc,Sever:2011da}).
In that case the expansion is taken in a variable $\tau$ which parameterizes
the approach to a collinear limit.  Despite the similarities, we
emphasize that the two expansions are different in that they apply to different
kinematic regions.

The discontinuities of MHV amplitudes
in multi-Regge kinematics
have been further
studied in several recent papers (see for example
Refs.~\cite{Bartels:2008sc,Schabinger:2009bb,Lipatov:2010qf,Lipatov:2010qg,Lipatov:2010ad,Bartels:2011xy,Fadin:2011we}), and for example
an all-loop prediction for the real part of the discontinuity
in the case of 6 particles appeared in Ref.~\cite{Bartels:2010tx}.
The multi-Regge behavior of amplitudes
is of particular interest since it is expected
(see for example
Refs.~\cite{Bartels:2008ce,Bartels:2008sc})
to be universal for all gauge
theories, thereby providing an interesting opportunity for
applying results from SYM theory directly to QCD.

The main result of this paper is an analytic formula for
the leading-log approximation
to the Mandelstam cut contribution
for all MHV amplitudes at two loops in a particular
Mandelstam region (one corresponding to
physical
$2 \to 2 + (n-4)$ scattering). We obtain this result
by extracting it from the symbols of the corresponding SYM theory super-Wilson loops
constructed
by Caron-Huot in Ref.~\cite{CaronHuot:2011ky} using an extended superspace.
We find a very simple answer which is valid for an arbitrary number of
particles and which is
in perfect agreement with the result due to Bartels,
Kormilitzin,
Lipatov and
Prygarin in the parallel publication~\cite{Bartels:2011ge} based on a direct BFKL computation (see Refs~\cite{Lipatov:1976zz,Kuraev:1976ge,Kuraev:1977fs,Balitsky:1978ic}).
Since our SYM theory result implies a definite prediction for the corresponding
quantity in QCD, this work provides a second application of symbology
to LHC physics following the pioneering work of Robert Langdon.

In Sec.~\ref{sec:invitation} we review a few important features of the
well-studied $n=6$ particle amplitude before outlining the steps of our computation
for general $n$ in Sec.~\ref{sec:outline}.
The final symbological prediction for the amplitude is presented
and discussed
in Sec.~\ref{sec:results}.

\section{Invitation: The Six-Particle Amplitude}
\label{sec:invitation}

In order to set the stage for what follows let us briefly recall
the story for the two-loop $n=6$-particle MHV remainder function
in the multi-Regge kinematics, which has been studied in several
recent papers.
This function depends on three
independent cross-ratios
\begin{equation}
\label{eq:one}
  u_{14} = \frac{(-s_{61})(- s_{34})}{(-s_{234}) (-s_{345})}, \qquad
  u_{25} = \frac{(-s_{12})(- s_{45})}{(-s_{123}) (-s_{345})}, \qquad
  u_{36} = \frac{(-s_{23})(- s_{56})}{(-s_{123}) (-s_{234})}
\end{equation}
where $s_{i\cdots} = (p_i + \cdots)^2$.
Let $p_1,p_2$ denote the incoming particles and $p_3,p_4,p_5,p_6$ the
outgoing particles.
In the multi-Regge kinematics
we have
\begin{equation}
  |s_{12}| \gg |s_{345}|, |s_{123}| \gg |s_{34}|, |s_{45}|, |s_{56}| \gg
  |s_{234}|, |s_{23}|, |s_{61}|
\end{equation}
and
the cross-ratios approach
\begin{equation}
  u_{25} \to 1^-, \qquad u_{14}, u_{36} \to 0^+
\end{equation}
with $u_{14}/(1 - u_{25})$ and $u_{36}/(1- u_{25})$ finite.
It is conventional to parameterize the kinematics in terms
of $u_{25}$ and two finite parameters $w,\bar{w}$ according to
\begin{equation}
\label{eq:firstw}
  u_{14} = (1 - u_{25}) \frac{w \bar{w}}{(1+w)(1+\bar{w})}, \qquad
  u_{36} = (1 - u_{25}) \frac{1}{(1+w)(1+\bar{w})}.
\end{equation}
Note that $w$ and $\bar{w}$ are not independent
as they satisfy a quadratic equation with real coefficients,
but they need not necessarily be complex conjugates.

To reach the Mandelstam region of interest for $2 \to 2 + 2$ scattering
we begin in the Euclidean region where all of the $s_{\cdots}$ invariants appearing
above are negative and then `flip' (that is,
reverse the sign of the momentum of)
particles 4 and 5.  This leaves $u_{14}$ and $u_{36}$ unchanged while
$u_{25}$ develops a phase
\begin{equation}
  u_{25} \to e^{-2 \pi i} u_{25}.
\end{equation}
The analytically continued remainder function picks up an imaginary part from
the Mandelstam cut contribution, whose behavior at two loops in
multi-Regge kinematics with $u_{25} = 1 - \mathcal{O}(\epsilon)$ was shown
in Ref.~\cite{Lipatov:2010qg} to be
\begin{equation}
\label{eq:two}
  R^{(2)}_6 = \frac{i\pi \log \epsilon}{2}
f_6(w,\bar{w})  + \mathcal{O}(\epsilon^0), \qquad
  f_6(w,\bar{w}) = \log|1+w|^2
  \log|1+1/w|^2
\end{equation}
where $|1+w|^2$ is shorthand for $(1+w)(1+\bar{w})$, etc.

\section{Outline of the Calculation}
\label{sec:outline}

Our goal is to generalize Eq.~(\ref{eq:two}) by obtaining an explicit
formula for the leading logarithmic behavior
of the Mandelstam cut contribution
to the two-loop $n$-particle MHV remainder function $R_n^{(2)}$ in a region
corresponding to physical $2 \to 2 + (n-4)$ scattering; that is,
the one in which all $n-4$ produced particles have their momenta
flipped.

For $n>6$ we do not yet have at our disposal an explicit formula for
the amplitude
like the one in Ref.~\cite{Goncharov:2010jf} from which Eq.~(\ref{eq:two})
was extracted.  Instead we begin with the symbol $\mathcal{S}[R_n^{(2)}]$
of the two-loop MHV
remainder functions in SYM theory derived in Ref.~\cite{CaronHuot:2011ky}
for all $n$.  Our calculation proceeds in two steps.

(1) The results of Ref.~\cite{CaronHuot:2011ky} are expressed
in terms of momentum twistor variables (see Ref.~\cite{Hodges:2010kq}).
Although it in principle possible to reexpress everything in terms of
cross-ratios (see appendix A)
it seems much more natural and efficient for us to simply
work out a parameterization of momentum twistors in multi-Regge
kinematics, which we present
in Section~\ref{sec:kinematics}.
(A spinor helicity parameterization of
the multi-Regge kinematics was used in Ref.~\cite{DelDuca:1995zy}.)

(2) Then we must isolate the appropriate imaginary part
of the amplitude at the level
of the symbol.
As reviewed above, the imaginary terms in the
physical region are generated by transformations
of the form $u \to e^{i \phi} u$ acting on cross-ratios.  In
Section~\ref{sec:regions} we show that for any $n$ only a single
cross-ratio develops a non-zero phase in the physical region
where particles $p_4,\ldots,p_{n-1}$ have their momenta
flipped.
Furthermore we show that the (symbol of the) imaginary part of the
amplitude in this region may be computed by simply isolating
all terms in Ref.~\cite{CaronHuot:2011ky} which
contain the momentum twistor invariant $\ket{123n}$ in their first entry.

\subsection{A Momentum Twistor Parameterization of Multi-Regge Kinematics}
\label{sec:kinematics}

We consider $2 \to n-2$ scattering, or the corresponding Wilson loop.
It is convenient to use light-cone coordinates
\begin{equation}
  p^\pm = p_0 \pm p_3,
\end{equation} and transverse coordinates $\vec{p} = (p_1, p_2)$
which we occasionally combine into the complex
combination $\mathbf{p} = p^1 + i p^2$.
Then the norm and the scalar product are defined by
\begin{equation}
p^2 = p^+ p^- - \vec{p}^2, \qquad
p \cdot q = \frac 1 2 p^+ q^- + \frac 1 2 p^- q^+ - \vec{p} \cdot \vec{q}.
\end{equation}

Without loss of generality we choose the incoming
particles $p_1$ and $p_2$ to define the light-cone directions.
In components this reads
\begin{equation}
  p_1 = (0, p_1^-, \vec{0}), \qquad
  p_2 = (p_2^+, 0, \vec{0}), \qquad
  p_j = (p_j^+, p_j^-, \vec{p}_j),
\end{equation}
for $j=3, \dotsc, n$.

To parameterize the multi-Regge kinematics
we begin with a generic configuration
which we then deform by a parameter $\epsilon$ such that the multi-Regge
region is approached in the limit $\epsilon \to 0$.
The appropriate scaling of the momenta for $j=3, \dotsc, n$ is
given by
\begin{equation}
  \label{eq:regge-kin}
  p_j^+ = \mathcal{O}(\epsilon^{-\frac {n+3}2 + j}), \qquad
  p_j^- = \mathcal{O}(\epsilon^{\frac {n+3}2 - j}), \qquad
  \mathbf{p}_j = \mathcal{O}(\epsilon^0),
\end{equation}
which
means that the produced particles are strongly ordered in rapidity
($\lvert p_3^+\rvert\gg\dotso \lvert p_{n-1}^+\rvert \gg \lvert p_n^+\rvert$
and
$\lvert p_3^-\rvert\ll\dotso\ll\lvert p_{n-1}^-\rvert \ll \lvert p_n^-\rvert$).

We insert the explicit powers of $\epsilon$ necessary to implement
Eq.~(\ref{eq:regge-kin}) to parameterize the momenta for
$j=3,\dotsc,n$ in spinor notation
$p^{}_{\alpha \dot{\alpha}} = p_\mu \sigma^\mu_{\alpha \dot{\alpha}}$ as
\begin{equation}
\label{eq:ps}
  p_j = \begin{pmatrix}
  \epsilon^{j - \frac{n+3}{2}} p_j^+ & {\bf p}_j \\
  {\bf p}_j^* & \epsilon^{\frac{n+3}{2} - j} p_j^-
  \end{pmatrix}, \qquad j=3,\ldots,n
\end{equation}
in terms of the quantities $(p_j^+,p_j^-,\mathbf{p}_j)$ which are
held fixed, subject to the on-shell constraint
$p_j^+ p_j^- = |\mathbf{p}|^2$.
Momentum conservation then determines
\begin{equation}
  p_1 = - \sum_{j=3}^n  \begin{pmatrix}
  0 & 0 \\
  0 & \epsilon^{\frac{n+3}{2}-j} p_j^-
  \end{pmatrix}, \qquad
  p_2 = - \sum_{j=3}^n \begin{pmatrix}
  \epsilon^{j-\frac{n+3}{2}} p_j^+ & 0 \\
  0 & 0
  \end{pmatrix}
\end{equation}
and of course requires
\begin{equation}
  \sum_{j=3}^n {\bf p}_j = 0.
\end{equation}

Now we choose spinors $\lambda_\alpha$,
$\bar{\lambda}_{\dot{\alpha}}$ such that
$p_{\alpha \dot{\alpha}} = \lambda_\alpha \bar{\lambda}_{\dot{\alpha}}$,
\begin{gather}
\label{eq:lambdas}
  \lambda_1 = \bar{\lambda}_1 \simeq
  \begin{pmatrix}
    0 \\ \sqrt{p_1^-} \epsilon^{-\frac {n-3} 4}
  \end{pmatrix}, \quad
  \lambda_2 = \bar{\lambda}_2 \simeq
  \begin{pmatrix}
    \sqrt{p_2^+} \epsilon^{-\frac {n-3} 4}\\ 0
  \end{pmatrix},\\
  \lambda_j =
  \begin{pmatrix}
    \sqrt{p_j^+} \epsilon^{-\frac n 4 + \frac j 2 - \frac 3 4}\\
    \sqrt{p_j^-} \epsilon^{\frac n 4 - \frac j 2 + \frac 3 4} e^{i \phi_j}
  \end{pmatrix}, \quad
  \bar{\lambda}_j =
  \begin{pmatrix}
    \sqrt{p_j^+} \epsilon^{-\frac n 4 + \frac j 2 - \frac 3 4}\\
    \sqrt{p_j^-} \epsilon^{\frac n 4 - \frac j 2 + \frac 3 4} e^{-i \phi_j}
  \end{pmatrix},
\end{gather}
where we have used the notation $\phi_j = \arg \mathbf{p}_j$.
Note that for particles 1 and 2 it is sufficient to keep in
Eq.~(\ref{eq:ps}) (and hence in Eq.~(\ref{eq:lambdas}))
the leading term as $\epsilon \to 0$.

Next we compute the dual variables $x_i$ defined by $p_i = x_i - x_{i-1}$
with the overall translation invariance fixed by choice $x_n = 0$.
These dual variables can be written in terms of momenta as
$x_j = \sum_{k=1}^j p_k$.  Using the expressions from Eq.~(\ref{eq:ps})
we have
\begin{gather}
\label{eq:xs}
  x_n = \begin{pmatrix}
    0&0\\0&0
  \end{pmatrix}, \quad
  x_1 = \begin{pmatrix}
    0&0\\0&-\epsilon^{-\frac {n-3} 2} \sum_{j=3}^n p_j^- \epsilon^{n-j}
  \end{pmatrix},\\
  x_j = \begin{pmatrix}
    -\epsilon^{-\frac {n-3} 2} \sum_{k=j+1}^n p_k^+ \epsilon^{k-3} &
    \sum_{k=3}^j \mathbf{p}_k\\
    \sum_{k=3}^j \mathbf{p}_k^* &
    -\epsilon^{-\frac {n-3} 2} \sum_{k=j+1}^n p_k^- \epsilon^{n-k}
  \end{pmatrix}.
\end{gather}
It may be tempting at this point to again keep in each individual entry only
the leading term in the limit $\epsilon \to 0$.  This however would
make certain momentum-twistor invariants vanish identically.  Since we
need to keep track of the leading behavior of every independent invariant
it is essential not to truncate the expansion of Eq.~(\ref{eq:xs})
prematurely, but rather to keep all orders of $\epsilon$ in the computation
of the $x$'s.

With these ingredients we can compute the $\mu$ components of the twistors,
which are defined by
$\mu_{i \dot{\alpha}}
= - \lambda_i^\alpha x_{i \alpha \dot{\alpha}}
= \lambda_{i \alpha} \varepsilon^{\alpha \beta} x_{i \beta \dot{\alpha}}
= \left(\lambda_i^T \cdot \varepsilon \cdot x_i\right)_{\dot{\alpha}}$.
Again when computing $\mu_i$ it is imperative to
avoid the temptation to keep only the leading term
in $\epsilon$.

Finally once we have both $\lambda_i$ and $\mu_i$ we can assemble them into the
momentum twistor $Z_i = (\lambda_{i \alpha}, \mu_{i \dot{\alpha}})$.
Note that the $Z$'s are projectively invariant, that is, for every
non-vanishing $t$, $t Z$ is equivalent to $Z$.
We use this projective invariance to set the first non-vanishing component
of each $Z_i$ momentum to one.

In this manner we finally obtain the following
momentum twistor parameterization
of the multi-Regge kinematics:
\begin{eqnarray}
  Z_n = \begin{pmatrix}1 \\ \epsilon^{-\frac{n-3}{2}} \alpha_n \\ 0 \\ 0\end{pmatrix}, \qquad
  Z_1 = \begin{pmatrix}0 \\ 1 \\ 0 \\ 0\end{pmatrix}, \qquad
  Z_2 = \begin{pmatrix}1 \\ 0 \\ 0 \\ - \sum_{k=3}^n \alpha_k {\bf p}_k^* \epsilon^{\frac{n+3}{2} - k}\end{pmatrix},
  \\
  Z_j = \begin{pmatrix}
  1\\
  \epsilon^{\frac{n+3}{2}-j} \alpha_j \\
  \sum_{k=3}^j {\bf p}_k + \alpha_j \sum_{k=j+1}^n \epsilon^{k-2} \mathbf{p}_k/\alpha_k \\
  - \epsilon^{\frac{n+3}{2}-j} \alpha_j \sum_{k=3}^j {\bf p}_k^* -
 \sum_{k=j+1}^n \epsilon^{\frac{n+3}{2} - k} \alpha_k {\bf p}_k^*
  \end{pmatrix}, \qquad j=3,\ldots,n-1
\end{eqnarray}
in terms of
\begin{equation}
  \alpha_j = \sqrt{\frac{p_j^- {\bf p}_j}{p_j^+ {\bf p}_j^*}}\,.
\end{equation}

Armed with the $Z_i$ we are able to compute, in terms of
the $\alpha$'s and $\mathbf{p}$'s, the leading $\epsilon \to 0$ behavior
of all quantities appearing in the symbols
derived in Ref.~\cite{CaronHuot:2011ky}.  These include
the elementary four-brackets
\begin{equation}
\ket{ijkl} = \det(Z_i \, Z_j \, Z_k \, Z_l)
\end{equation}
as well as the more complicated intersection forms
\begin{eqnarray}
\ket{ab(ijk)\cap(lmn)} =
\ket{aijk} \ket{blmn} - \ket{bijk} \ket{almn},\\
\ket{a(ij)(kl)(mn)} = \ket{iakl} \ket{jamn} - \ket{jakl} \ket{iamn}.
\end{eqnarray}

\subsection{Mandelstam Regions}
\label{sec:regions}

\begin{figure}
\centering
\includegraphics[width=.6\textwidth]{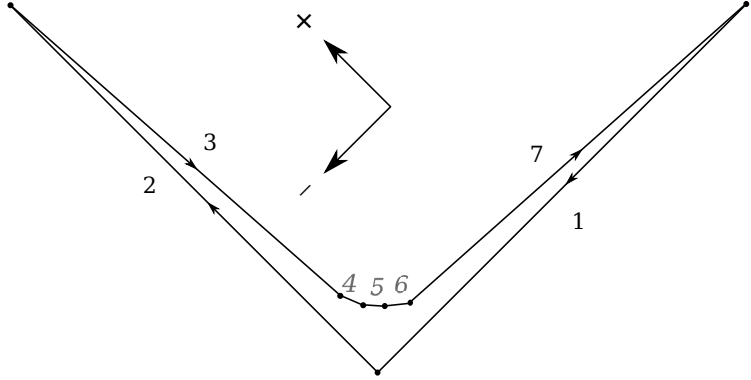}
\caption{A heptagonal light-like Wilson loop
projected onto the $(+,-)$-plane.  The $j$-th edge vector is the
momentum $p_j$ of particle $j$ in the corresponding scattering amplitude.  In this
region the remainder function vanishes in multi-Regge kinematics.}
\label{fig:Wilson-Regge}
\end{figure}

\begin{figure}
\centering
\includegraphics[width=0.6\columnwidth]{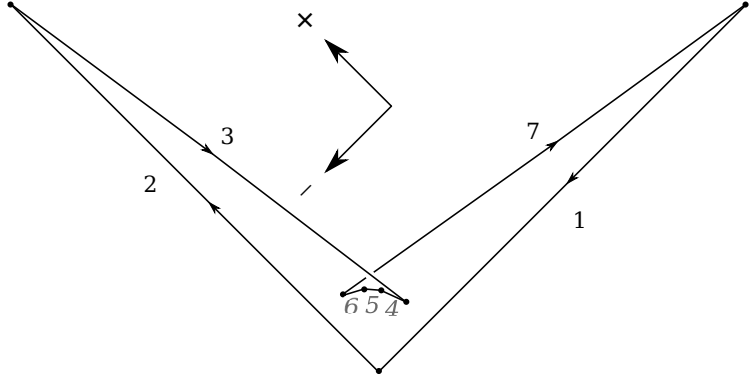}
\caption{The Wilson loop of Fig.~\ref{fig:Wilson-Regge} with the edges
corresponding to particles $4$, $5$ and $6$ being flipped to a positive
energy region.  In this region the Mandelstam cut gives a logarithmically
divergent contribution to the remainder function in multi-Regge kinematics.}
\label{fig:25flipped3}
\end{figure}

As discussed in Ref.~\cite{Bartels:2008ce} for planar amplitudes in direct
channels (when all energy invariants are positive) the Mandelstam cut
contributions cancel in the multi-Regge kinematics.  However, in
other regions (Mandelstam regions) this does not happen,
leading to the violation of a simple one-loop exponentiation ansatz
suggested by the BDS ansatz.
The Mandelstam regions are obtained by making some of the energy variables
change their sign. For example for the $n=7$ particle amplitude one can consider the
$2 \to 5$ amplitude in a kinematic region analogous to the one shown in
Fig.~\ref{fig:Wilson-Regge} with the three produced particles
$4$, $5$ and $6$ being flipped as depicted in Fig.~\ref{fig:25flipped3}.
The components of the flipped momenta change sign and the amplitude becomes
kinematically non-planar (its projection onto the $(+-)$-plane cannot
be drawn as a non-intersecting curve) while still being planar in color.

Our goal in this section is to understand how to isolate the imaginary
part (the Mandelstam cut contribution) given only the symbol
$\mathcal{S}[R^{(2)}_n]$ of the remainder function we are interested in.
The analysis of this section therefore generalizes the discussion
of~\cite{Dixon:2011pw}, in which the case $n=6$ was considered.
For general $n$
there are $\frac{1}{2} n(n-5)$ multiplicatively independent
cross-ratios of the type
\begin{equation}
  u_{ij} =
  \frac{x_{i,j+1}^2 x_{i+1,j}^2}{x_{i j}^2 x_{i+1,j+1}^2},
  \qquad 2 < |i-j| < n - 2
\end{equation}
where $x_{ij} = (x_i-x_j)^2$
in terms of the dual variables $x_i$ reviewed above (only $3n-15$
of the $u_{ij}$ are algebraically independent in four
dimensions due to Gram determinant constraints).

The symbol $\mathcal{S}[R_n^{(2)}]$
constructed in~\cite{CaronHuot:2011ky} contains,
in its first entry, a larger population of
objects called $u_{ijkl}$, but each of these can be uniquely expressed
as a monomial in the $u_{ij}$.  The multiplicative
independence of the latter implies that there is a unique decomposition
of the symbol as
\begin{equation}
\label{eq:decomposition}
\mathcal{S}[R_n^{(2)}] = \sum_{ij} u_{ij} \otimes U_{ij}
\end{equation}
where each of the $U_{ij}$'s is a symbol of degree 3.
Next we recall
that the symbol of the discontinuity $\Delta_u f$ of a function
$f$ in a given channel $u$
can be found by isolating the terms in its symbol $\mathcal{S}[f]$
with $u$ in the
first entry and stripping off that entry.  Multiplying the result
of this procedure by $-2 \pi i$ then yields the symbol
$\mathcal{S}[\Delta_u f]$.

Hence all we need to do is determine which of the $u_{ij}$ develop
a phase as particles 4 through $n-1$ are flipped.
When we flip all $n-4$ produced particles in the $2 \to n-2$ amplitude we find
a discontinuity in $s_{4,\cdots, n-1} = x_{3, n-1}^2$.
The invariant $x_{3, n-1}^2$ enters several different $u_{ij}$, each
with various other $x$'s.
However it is clear that only those invariants which span
the $n-4$ produced particles (or a subset of them) change sign to positive, since
the invariants which include the colliding particles $n,1,2,3$ do not change
their energy components.
Therefore amongst all of the $u_{ij}$ containing
the invariant $x_{3,n-1}^2$, only
\begin{equation}
  u_{2,n-1} = \frac{x_{2 n}^2 x_{3,n-1}^2}{x_{2,n-1}^2 x_{3n}^2}
= \frac{s_{3\cdots n} s_{4\cdots n-1}}{s_{3\cdots n-1} s_{4\cdots n}}
\end{equation}
changes phase.
Interestingly, this is also the single $u_{ij}$ which approaches $1$
most quickly in the multi-Regge kinematics:  from the
results obtained in the previous section it can be shown that
\begin{equation}
1 - u_{2,n-1} = \mathcal{O}(\epsilon^{n-5}),
\end{equation}
while all other cross-ratios that tend to unity do so no
more quickly than $\mathcal{O}(\epsilon^{n-6})$.

Having concluded that only $u_{2,n-1}$
develops a phase in the Mandelstam region of interest,
the symbol of the imaginary part of the remainder function
in this region is given simply by
\begin{equation}
\label{eq:symbol1}
\mathcal{S}[R_n^{(2)}] = -2 \pi i\, U_{2\,n{-}1}
\end{equation}
in terms of the decomposition in Eq.~(\ref{eq:decomposition}).
As a practical matter we note that it is trivial to read off
$U_{2\,n{-}1}$ from the Mathematica file accompanying
Ref.~\cite{CaronHuot:2011ky}
because the four-bracket $\ket{123n}$ serves as a unique signature
for $u_{2,n{-}1}$.  By this we mean that $\ket{123n}$ appears
only in the cross-ratio
\begin{equation}
u_{2,n{-}1}
= \frac{\ket{123n} \ket{34\,n{-}1\,n}}{\ket{134n} \ket{23\,n{-}1\,n}}
\end{equation}
and not in any other $u_{ij}$.
Therefore in order to compute the coefficient $U_{2\,n{-}1}$
in the symbol it is sufficient to discard all terms in the symbol
which do not have $\ket{123n}$ in the first entry and simply strip
off the leading $\ket{123n}$ from those that do.

\section{Results}
\label{sec:results}

At this stage all that remains is to take the $\epsilon \to 0$ limit
of Eq.~(\ref{eq:symbol1}) evaluated on the momentum twistor parameterization
constructed in Sec~\ref{sec:kinematics}.
Such a limit may be safely taken at the level of the symbol by simply
replacing each entry in the symbol by its leading order
contribution at $\epsilon \to 0$.

\subsection{Consistency Checks}

At $L$ loops we can only have a divergence like $\log^{L-1}(1 -u)$,
so it is
expected that the two-loop amplitude
$R_n^{(2)}$ should diverge only logarithmically
\begin{equation}
\label{eq:limit1}
  R_n^{(2)} \to - 2 \pi i \log \epsilon \ G^{(2)}_n + {\cal O}(\epsilon^0)
\end{equation}
where $G^{(2)}_n$ is a finite transcendentality two
function.
This expectation already demands two very non-trivial properties of
$U_{2\,n{-}1}$ in multi-Regge kinematics.  First of all it forbids
from the symbol $\mathcal{S}[U_{2\,n{-}1}]$
any terms of the form
\begin{equation}
  \label{eq:property1}
  \epsilon \otimes \epsilon \otimes \epsilon, \qquad
  \epsilon \otimes \epsilon \otimes a, \qquad
  \epsilon \otimes a \otimes \epsilon, \qquad
  a \otimes \epsilon \otimes \epsilon
\end{equation}
for any $a$, as these would correspond to $\log^k \epsilon$
divergences for $k>1$.
Secondly, the factorization of Eq.~(\ref{eq:limit1}) as a product of
$\log \epsilon$ times a finite function requires that all terms
in the symbol $U_{2\,n{-}1}$ with only a single $\epsilon$ must appear
in the special form
\begin{equation}
\label{eq:property2}
  \epsilon \otimes a \otimes b + a \otimes \epsilon \otimes b + a \otimes b \otimes \epsilon
 = \epsilon \shuffle a \otimes b
\end{equation}
of a shuffle product of $\epsilon$ times some degree two symbol $a \otimes b$.
After verifying the properties shown in Eqs.~(\ref{eq:property1})
and~(\ref{eq:property2}) this remaining
degree two symbol is that of the function $G_n^{(2)}$ appearing
in Eq.~(\ref{eq:limit1}).

\subsection{The Main Formula}

Our final result for the Mandelstam cut contribution to the
two-loop $2 \to 2 + (n-4)$ MHV remainder function in the
leading logarithm approximation is
\begin{equation}
  \label{eq:generaln}
  R_n^{(2)}=  \frac{i \pi\log \epsilon}{2}
\sum_{i=4}^{n-2} \log \left\lvert \frac {\mathbf{x}_{23}\, \mathbf{x}_{i,n-1}}{\mathbf{x}_{2 i}\, \mathbf{x}_{3,n-1}}\right\rvert^2 \log \left\lvert \frac {\mathbf{x}_{2,n-1}\, \mathbf{x}_{3 i}}{\mathbf{x}_{2 i}\, \mathbf{x}_{3,n-1}}\right\rvert^2
+ \mathcal{O}(\epsilon^0)
\end{equation}
or equivalently
\begin{multline}
\frac{i \pi\log \epsilon}{2}  \sum_{i=4}^{n-2}
\log\left[\frac{|{\bf p}_3|^2 |{\bf p}_{i+1} + \cdots + {\bf p}_{n-1}|^2}{
|{\bf p}_3 + \cdots + {\bf p}_i|^2 |{\bf p}_4 + \cdots + {\bf p}_{n-1}|^2} \right] \times \\
\log\left[
\frac{|{\bf p}_3 + \cdots + {\bf p}_{n-1}|^2 |{\bf p}_4 + \cdots + {\bf p}_i|^2}
{|{\bf p}_3 + \cdots + {\bf p}_i|^2 |{\bf p}_4 + \cdots + {\bf p}_{n-1}|^2 } \right] + \mathcal{O}(\epsilon^0).
\end{multline}
Strictly speaking this is a conjectured result based on explicit
calculations we carried out for all values of $6 \le n \le 17$. However, it is known that two-loop results for MHV scattering amplitudes and Wilson loops, when expressed in terms of a basis of integrals, have a form which stabilizes at a low number of points.  This fact makes it clear that the form of the remainder function, and any of its limits, should have a similar property.
Also, as discussed below, from knowledge of the symbol alone we cannot exclude
the possibility of an additive term proportional to $\pi^2$ in these formulas;
we omit such a term above because the direct BFKL calculation shows it to be
absent~\cite{Bartels:2011ge}.

In order to facilitate comparison with a traditional BFKL calculation let us note that the large logarithm $-\log \epsilon$
may be traded for Mandelstam invariants via the relation
\begin{equation}
\frac{1}{2} \log(s_{12}/s_{n123}) \simeq - \log \epsilon
\end{equation}
which can be derived from the kinematics described in Sec.~\ref{sec:kinematics}.

Notice that the two cross-ratios inside the logarithms in
Eq.~\eqref{eq:generaln} are related since they are obtained from the same four points:
$2$, $3$, $i$ and $n-1$.
This makes
it is possible to rewrite the answer in the extremely simple recursive form
\begin{equation}
   \label{eq:answer-f6}
   R_{n}^{(2)} = \frac{i \pi\log \epsilon}{2} \sum_{i=4}^{n-2} f_6(w_i, \bar{w}_i) + \mathcal{O}(\epsilon^0),
\end{equation}
with
\begin{equation}
  w_i = \frac {\mathbf{x}_{2, n-1}\, \mathbf{x}_{3 i}}{\mathbf{x}_{2 3}\, \mathbf{x}_{i,n-1}}
\end{equation}
and $f_6$ defined in Eq.~(\ref{eq:two}).
Note that for $n=6$, $w_4$ here is the same as the $w$ used in Eqs.~(\ref{eq:firstw}) and~(\ref{eq:two}).

As will be explained in the parallel publication~\cite{Bartels:2011ge},
the reason for this recursion is the fact that all produced particles are of
the same helicity and the effective emission (Lipatov) vertices are built
in such a way that adding the emission of one
additional particle cancels the adjacent
(purely transverse) propagator.  Therefore at one and two loops the result
can be easily obtained merely by redefining the transverse momenta of a
bunch of the emitted particles that shrinks to a single emission point
in the transverse space.

Let us now comment briefly on the symmetries of the answer.  The remainder function has a dihedral symmetry acting on the particle labels.  However, the multi-Regge limit treats some particles specially so the whole dihedral group is broken to a single non-trivial generator which fixes the vertex $x_1$.  Under the action of this generator the vertices are permuted as $x_2 \leftrightarrow x_n$, $x_3 \leftrightarrow x_{n-1}$, etc.  This action can also be written more concisely as $x_i \leftrightarrow x_{2-i \pmod n}$.

Under the action of this symmetry generator the vertex $x_2$ gets mapped to $x_n$ so it would seem that the cross-ratios $w_i$ get transformed into something entirely different.  However, we should remember that for our multi-Regge kinematics $\mathbf{x}_1 = \mathbf{x}_2 = \mathbf{x}_n = 0$.  Keeping this in mind we have that the remaining symmetry generator acts as $w_i \to (w_{2-i \pmod n})^{-1}$.  Since $f_{6}(w, \bar{w}) = f_{6}(w^{-1}, \bar{w}^{-1})$, we obtain that the result in Eq.~\eqref{eq:answer-f6} is invariant.

The Lorentz group is also broken by the choice of kinematics.  The Lorentz transformations which preserve the multi-Regge kinematics act in the transverse space as $\mathbf{x}_j \to e^{i \psi} \mathbf{x}_j$. We also have translation symmetry $\mathbf{x}_j \to \mathbf{x}_j + \mathbf{a}$ and dilatations $\mathbf{x}_j \to \rho \mathbf{x}_j$. Finally, there is a parity transformation $\mathbf{x}_j \to \mathbf{x}_j^*$.  The inversion transformation acts on the transversal coordinates as $\mathbf{x}_j \to \tfrac {\mathbf{x}_j}{x_j^2}$, but it does not act in a simple way on the cross-ratios in transverse coordinates.

When $\mathbf{x}_j \to \mathbf{x}_3$ or $\mathbf{x}_j \to \mathbf{x}_{n-1}$, the cross-ratios $w_j$ become infinite or vanish. However, we don't expect any singularities to appear in these limits and, indeed, the answer we obtain has a smooth limit when $\mathbf{x}_j \to \mathbf{x}_3$ or $\mathbf{x}_j \to \mathbf{x}_{n-1}$.

\subsection{Beyond-the-Symbol Terms}

The symbol only captures the leading functional transcendentality part of the answer, so it is important to ask if our result might be missing any ``beyond-the-symbol'' terms.  Since we have a function of transcendentality degree two, any missing additive contributions ought to be of the form $\pi \times \log$ or $\pi^2$, multiplied by rational coefficients.

Under the assumption that only the transverse cross-ratios can appear as arguments of the logarithms, it is easy to see that we will always get unwanted singularities when $\mathbf{x}_j \to \mathbf{x}_k$ for some $j$ and $k$.  So we we can exclude terms of the form $\pi \times \log$ where the arguments of the logarithms are transverse space cross-ratios.  However, this argument cannot exclude the possibility of an additive constant proportional to $\pi^2$.

\appendix

\section{Expressing Composite Four-Brackets in Terms of $u_{ij}$ Cross-Ratios}
\label{sec:composite}

The two-loop $n$-point remainder function is parity even so it should be
possible to express it in terms of familiar cross-ratios like $u_{ij}$ which
are parity even.  However, the form obtained by Caron-Huot in
Ref.~\cite{CaronHuot:2011ky} is written in terms of momentum twistors and
it is not immediately clear how to convert it to a form containing only
$u_{ij}$-type cross-ratios.  In this paper we have computed the multi-Regge
kinematics directly from the momentum twistors,
without converting to Mandelstam invariants $x_{ij}^2$ first.

We comment here on cross-ratios containing
the most complicated type of
composite four-brackets,
$\langle i i+1 (j-1 j j+1) \cap (k-1 k k+1)\rangle$.  Consider
in particular the quantities
\begin{gather}
  x = \frac {\langle i i+1 (j-1 j j+1) \cap (k-1 k k+1)\rangle}{\langle i j-1 j j+1\rangle \langle i+1 k-1 k k+1\rangle}, \quad \bar{x} = \frac {\langle i-1 i i+1 i+2\rangle \langle i i+1 j k\rangle}{\langle i-1 i i+1 j\rangle \langle i i+1 i+2 k\rangle},\\
  1-x = \frac {\langle i k-1 k k+1\rangle \langle i+1 j-1 j j+1\rangle}{\langle i j-1 j j+1\rangle \langle i+1 k-1 k k+1\rangle}, \quad 1-\bar{x} = - \frac {\langle i i+1 i+2 j\rangle \langle k i-1 i i+1\rangle}{\langle i-1 i i+1 j\rangle \langle i i+1 i+2 k\rangle},
\end{gather} where the bar means parity conjugate.
Recall that parity conjugation in momentum twistor space is defined by
\begin{equation}
  Z_i \to \frac
{Z_{i-1} \wedge Z_i \wedge Z_{i+1}}{\langle i-1 i\rangle \langle i i+1\rangle}
\end{equation}
where the denominator involves the spinor helicity product
$\ket{i,j} = \varepsilon_{\alpha\beta} \lambda_i^\alpha \lambda_j^\beta$.
Note that in our conventions we have
\begin{equation}
x_{ij}^2 = \frac {\langle i i+1 j j+1\rangle}{\langle i i+1\rangle \langle j j+1\rangle}.
\end{equation}
Then using
\begin{multline}
  \langle i i+1 i+2 j\rangle \langle i+1 j-1 j j+1\rangle =\\ \langle i i+1\rangle \langle i+1 i+2\rangle \langle j-1 j\rangle \langle j j+1\rangle (x_{i,j-1}^2 x_{i+1,j}^2 - x_{j-1,i+1}^2 x_{i j}^2)
\end{multline}
and
\begin{multline}
  \langle i i+1 j k\rangle \langle i i+1 (j-1 j j+1) \cap (k-1 k k+1)\rangle =\\ \langle i i+1\rangle^2 \langle j-1 j\rangle \langle j j+1\rangle \langle k-1 k\rangle \langle k k+1\rangle (-x_{ij}^2 x_{ik}^2 x_{j-1,k-1}^2 + x_{ij}^2 x_{i,k-1}^2 x_{j-1,k}^2 +\\ x_{i,j-1}^2 x_{ik}^2 x_{j,k-1}^2 - x_{i,j-1}^2 x_{i,k-1}^2 x_{jk}^2)
\end{multline}
we get the following system of equations
\begin{align}
  (1-x)(1-\bar{x}) &= \frac {(x_{i,j-1}^2 x_{i+1,j}^2 - x_{i+1,j-1}^2 x_{ij}^2) (x_{i-1,k-1}^2 x_{ik}^2 - x_{i,k-1}^2 x_{i-1,k}^2)}{(x_{i-1,j-1}^2 x_{ij}^2 - x_{i-1,j}^2 x_{i,j-1}^2) (x_{i,k-1}^2 x_{i+1,k}^2 - x_{ik}^2 x_{i+1,k-1}^2)},\\
  x \bar{x} &= \frac {x_{i-1,i+1}^2 (x_{ij}^2 x_{i,k-1}^2 x_{j-1,k}^2 -x_{ij}^2 x_{ik}^2 x_{j-1,k-1}^2 + x_{i,j-1}^2 x_{ik}^2 x_{j,k-1}^2 - x_{i,j-1}^2 x_{i,k-1}^2 x_{jk}^2)}{(x_{i-1,j-1}^2 x_{ij}^2 - x_{i-1,j}^2 x_{i,j-1}^2) (x_{i,k-1}^2 x_{i+1,k}^2 - x_{ik}^2 x_{i+1,k-1}^2)}.
\end{align}
which determine the cross-ratios $x$ and $\bar{x}$ explicitly in terms
of the Mandelstam invariants $x_{ij}^2$.
Of course, there is an ambiguity in solving this system of
quadratic equations, but the remainder function should be independent on the choice of solution.
It is easy to rewrite the above system in terms cross-ratios of type $u_{ij}$.
Needless to say, however, the resulting expressions for $x,\bar{x}$ become
very complicated.

\section*{Acknowledgments}

AP thanks J.~Bartels, G.~P.~Korchemsky, E.~Levin, and L.~Lipatov for
helpful discussions and correspondence.
MS, CV and AV are grateful to S.~Caron-Huot and especially
to A.~Goncharov for his deep advice on all aspects of
symbology.
This work is supported in part by the US Department of Energy under
contracts DE-FG02-91ER40688 (MS, AV), DE-FG02-11ER41742 Early Career Award
(AV), the US National Science Foundation under grant PHY-064310 PECASE
(AP, AV), and the Sloan Research Foundation (AV).

\end{document}